\documentclass[a4paper]{jpconf}
\usepackage[psamsfonts]{amsfonts}
\usepackage{amssymb}
\usepackage{amsmath}
\usepackage{amsopn}
\usepackage{amsthm}
\usepackage{bm}
\usepackage{cite}
\usepackage{graphicx}

\DeclareMathOperator{\Integer}{\mathbb{Z}}

\DeclareMathOperator{\Ibb}{\mathbb{I}}

\DeclareMathOperator{\Mat}{Mat}

\DeclareMathOperator{\res}{res}
\DeclareMathOperator{\rank}{rank}

\theoremstyle{definition}
\newtheorem*{note*}{Note}

\theoremstyle{plain}
\newtheorem*{conjecture*}{Conjecture}
\newtheorem*{proposition*}{Proposition}

\allowdisplaybreaks[4]
\begin{document}
\title{SU(3) magnet: finite-gap integration \\ on the lowest genus curve}

\author{Julia Bernatska, Petro Holod}

\address{National University of `Kyiv Mohyla Academy', G. Skovorody St., 2,
  04655 Kyiv, UA}

\ead{BernatskaJM@ukma.kiev.ua, Holod@ukma.kiev.ua}

\begin{abstract}
We consider the integrable system of isotropic SU(3) Landau-Lifshits equation as a Hamiltonian
system on a coadjoint orbit of the SU(3) loop group.
We connect the mentioned equation with an isotropic SU(3) magnet because it describes 
the mean fields of magnetic and quadrupole moments in a spin-1 lattice. 
For the system of isotropic SU(3) Landau-Lifshits equation
we perform separation of variables in Sklyanin's manner,
and integrate in the lowest finite gap where the spectral curve is elliptic.

\end{abstract}

\section{Introduction}
The interest to integrable continuous spin chains starts from \cite{NakSas1974}, where
the isotropic Landau-Lifshitz equation
was considered as a spin equation of motion, soliton and periodic solutions were
obtained by the method of traveling wave.
Besides numerical investigations of soliton-soliton scattering and stability of soliton solutions
in \cite{LRT1976,TW1977},
the inverse scattering method was applied to the continuous Heisenberg spin chain in \cite{Takh1977}.
All the mentioned papers deal with SU(2) spin chain,
and the Landau-Lifshitz equation serves as an equation of motion.
A generalization to SU($n$) Heisenberg spin chain (Hamiltonian with bilinear exchange only) derived from the vector
nonlinear Sch\"{o}dinger equation was presented in \cite{Orf1980}.
In \cite{SaRuij1982} the integrable SU(3) spin chain with the most general Hamiltonian
was introduced as a system with the Lax-pair representation,
and shown to be gauge equivalent to a set of nonlinear Sch\"{o}dinger-type equations.
Separation of variables for the classical integrable SL(3)
magnetic chain was developed in \cite{Sklya92}.
In \cite{HKK1997} the isotropic SU(3) Landau-Lifshitz equation together
with two- and three-component nonlinear Sch\"{o}dinger-type equations were constructed by the orbit method.
In \cite{BernHol2009} the isotropic SU(3) Landau-Lifshitz equation is derived as a continuum limit
of the quantum spin-1 lattice system; this and a more general equations were constructed by means of the orbit method.
The latter generalization of the Landau-Lifshitz equation, obtained over a generic coadjoint orbit of SU(3),
is similar to the equation of motion for the SU(3) spin chain with the most general Hamiltonian from \cite{SaRuij1982}.

In the present paper we apply Sklyanin's separation of variables
to the isotropic SU(3) Landau-Lifshitz equation over a generic orbit.
In contrast to the classical integrable SL(3) magnetic chain from \cite{Sklya92},
whose phase space is endowed with a quadratic Poisson bracket and
the monodromy matrix serves as an $L$-operator,
we consider the system whose dynamic variables describe the mean fields of magnetic
and quadrupole moments and phase space is endowed with a Lie-Poisson bracket.
Nevertheless, we obtain similar expressions for variables of separation.
Then we proceed to finite-gap integration of the isotropic SU(3) Landau-Lifshitz equation.
According to \cite{Sklya95} a half of variables of separation serve as poles of
the Baker-Akhiezer function. For the system over a generic orbit
the number of these poles exceeds the genus of the spectral curve by two,
so integration leads to inversion of an extended Abel-Jacobi map by means of generalized $\theta$-functions
\cite{Fed1999,BraFed2008}.
The similar situation is considered in \cite{Adams}. We focus on the simplest one-gap case with
an elliptic spectral curve, and express Abelian integrals in terms of the Weierstrass elliptic function.

The structure of the paper is the following.
In Section 2 we describe quantum and classical models of SU(3) spin-1 chain.
The quantum spin operators are represented over
the 3-dimensional space of irreducible representation of the group SU(2).
We construct quadrupole operators from the tensor operators of weight 2.
These operators allow to linearize the system with the bilinear-biquadratic
exchange Hamiltonian. By the mean field approximation we take the continuum limit
and obtain a dynamical system coinciding with the one constructed over a degenerate coadjoint orbit
of SU(3). At the same time another dynamical system is constructed
over a generic coadjoint orbit of SU(3).
For the latter system in Section 3 we develop the scheme of separation of variables in Sklyanin's manner,
and prove that variables of separation are quasi-canonical.
We solve the problem of computing dynamic variables from the variables of separation.
Then we perform integration for the one-gap case in terms of the Weierstrass elliptic function.

\section{Quantum and classical models of SU(3) magnet}
\subsection{Quantum spin-1 lattice}
Suppose we have a lattice of atoms of spin 1.
Every atom is modeled by a tuple of spin and quadrupole operators.
In contrast to the well-known Heisenberg model considering two spin projections and representing spin operators
by the Pauli matrices, we take into account all three
projections of the spin momentum and represent spin operators over the 3-dimensional space:
\begin{gather*}
\hat{S}_x = \frac{1}{\sqrt{2}} \begin{pmatrix} 0&0&1\\
0&0&1\\ 1&1&0 \end{pmatrix},\quad
\hat{S}_y = \frac{1}{\sqrt{2}}  \begin{pmatrix} 0&0&-\rmi\\
0&0&\rmi \\ \rmi&-\rmi&0 \end{pmatrix},\quad
\hat{S}_z = \begin{pmatrix} 1&0&0\\
0&-1&0\\ 0&0&0 \end{pmatrix}.
\end{gather*}
The corresponding matrix algebra $\Mat_{3\times 3}$ is not exhausted by these three operators,
so we add six operators more.
We choose the following operators and call them quadrupole:
\begin{gather*}
\hat{P}_{xy} = \hat{S}_x \hat{S}_y + \hat{S}_y \hat{S}_x,\quad
\hat{P}_{xz} = \hat{S}_x \hat{S}_z + \hat{S}_z \hat{S}_x,\quad
\hat{P}_{yz} = \hat{S}_y \hat{S}_z + \hat{S}_z \hat{S}_y,\\
\hat{P}_{xx} = \hat{S}_x^2,\quad
\hat{P}_{yy} = \hat{S}_y^2,\quad
\hat{P}_{zz} = \hat{S}_z^2.
\end{gather*}

Introducing the quadrupole operators we have an opportunity to take into account
the second order of exchange interaction and use the bilinear-biquadratic exchange Hamiltonian, easily
reduced to a bilinear form:
\begin{align*}
\hat{\mathcal{H}} &= - \textstyle\sum_{n,\delta} \Big\{J \big(\hat{\bm{S}}_n,\hat{\bm{S}}_{n+\delta}\big)
+ K \big(\hat{\bm{S}}_n,\hat{\bm{S}}_{n+\delta}\big)^2 \Big\} \\
&= - (J-\tfrac{1}{2}\,K) \textstyle\sum_{n,\delta} \big(\hat{\bm{S}}_n,\hat{\bm{S}}_{n+\delta}\big)
- \tfrac{1}{2}\,K \sum_{n,\delta} \big(\hat{\bm{P}}_n,\hat{\bm{P}}_{n+\delta}\big) - \tfrac{4}{3} K N,
\end{align*}
where $J$ and $K$ are interaction constants, $n$ runs over all  $N$ sites of the lattice and
$\delta$ runs over the nearest-neighbour sites. We use the notation:
$\hat{\bm{S}}_n\,{=}\,\big(\hat{S}_x^{(n)}$, $\hat{S}_y^{(n)}$, $\hat{S}_z^{(n)}\big)$,
$\hat{\bm{P}}_n\,{=}\,\big(\hat{P}_{xy}^{(n)}$, $\hat{P}_{xz}^{(n)}$, $\hat{P}_{yz}^{(n)}$,
$\hat{P}_{xx}^{(n)}\,{-}\,\hat{P}_{yy}^{(n)}$, $\sqrt{3}(\hat{P}_{zz}^{(n)}\,{-}\,\frac{2}{3}\Ibb)\big)$, where
every entry $\hat{O}^{(n)}$ is a tensor product over all sites of $N\,{-}\,1$ identity operators $\Ibb$ and
the operator $\hat{O}$ at the $n$th position.

Next, we apply the mean field approximation under the broken symmetry stimulated by an external magnetic field.
This guarantees nonzero values of the quasiaverage fields corresponding
to the spin and quadrupole operators:
\begin{gather*}
\mu_1(x_n) = \langle \hat{S}_x^{(n)} \rangle,\quad \mu_2(x_n) = \langle \hat{S}_y^{(n)} \rangle, \quad
\mu_3(x_n) = \langle \hat{S}_z^{(n)} \rangle, \\
\tau_{12}(x_n) = \langle \hat{P}_{xy}^{(n)} \rangle,\quad
\tau_{13}(x_n) = \langle \hat{P}_{xz}^{(n)} \rangle,\quad
\tau_{23}(x_n) = \langle \hat{P}_{yz}^{(n)} \rangle,\\
\tau_{11}(x_n) = \langle \hat{P}_{xx}^{(n)} \rangle - \tfrac{2}{3},\quad
\tau_{22}(x_n) = \langle \hat{P}_{yy}^{(n)} \rangle - \tfrac{2}{3},\quad
\tau_{33}(x_n) = \langle \hat{P}_{zz}^{(n)} \rangle - \tfrac{2}{3}.
\end{gather*}
We arrange the mean fields into the magnetic moment vector-field $\bm{\mu}\,{=}\,(\mu_1,\mu_2,\mu_3)$,
and the quadrupole moment tensor-field $\hat{\tau}\,{=}\,(\tau_{ij})$, which is traceless.
In the mean field Hamiltonian the operators of neighbors in number $z$ are replaced by the corresponding mean fields:
\begin{align*}
\hat{\mathcal{H}}_{\text{MF}} &=
- (J-\tfrac{1}{2}\,K)z\sum_{n}\big(\hat{\bm{S}}_n,\langle \hat{\bm{S}}_n\rangle\big)
- \tfrac{1}{2}\,K z\sum_{n} \big(\hat{\bm{P}}_n,\langle\hat{\bm{P}}_n\rangle\big)- \tfrac{4}{3} K N z =\\
& = - (J-\tfrac{1}{2}\,K)z\sum_{n}\big(\mu_1\hat{S}_{x}^{(n)} + \mu_2\hat{S}_{y}^{(n)} + \mu_3 \hat{S}_{z}^{(n)}\big)
- \tfrac{1}{2}\,K z\sum_{n} \big(\tau_{12}\hat{P}_{xy}^{(n)} +
\\ &\phantom{=} + \tau_{13}\hat{P}_{xz}^{(n)}  + \tau_{23}\hat{P}_{yz}^{(n)}
+ (\tau_{11} - \tau_{22})\big(\hat{P}_{xx}^{(n)} - \hat{P}_{yy}^{(n)}\big) + 3\tau_{33}(\hat{P}_{zz}^{(n)}\,{-}\,\tfrac{2}{3}\Ibb)\big)
- \tfrac{4}{3} K N z.
\end{align*}

In what follows we consider the case of $K\,{=}\,J$,
when the Hamiltonian is uniform in the spin and quadrupole operators:
\begin{equation*}
\hat{\mathcal{H}}_{\text{MF}} =
\tfrac{1}{2}\,Jz \sum_n \sum_a \mu_a \hat{S}_a^{(n)} - \tfrac{4}{3} J N z.
\end{equation*}
We call this case isotropic SU(3) because
the operators $\{\hat{S}_x^{(n)}$, $\hat{S}_y^{(n)}$, $\hat{S}_z^{(n)}$, $\hat{P}_{xy}^{(n)}$, $
\hat{P}_{xz}^{(n)}$, $\hat{P}_{yz}^{(n)}$, $\hat{P}_{xx}^{(n)}\,{-}\,\hat{P}_{yy}^{(n)}$, $
\sqrt{3}(\hat{P}_{zz}^{(n)}\,{-}\,\tfrac{2}{3}\Ibb)\}$ form a basis of the algebra $\rmi\mathfrak{su}(3)$;
we denote them all together by $\{\hat{S}_a^{(n)}\}_{a=1}^8$, and the corresponding mean fields by
$\{\mu_a\}_{a=1}^8$, which are functions in the spatial coordinate $x$ and the time $t$.
For the mean field model we write Heisenberg's equations of motion
\begin{equation*}
\rmi \hbar \frac{d \hat{S}_a^{(n)}}{dt} = [\hat{S}_a^{(n)},\hat{\mathcal{H}}_{\text{MF}}],
\end{equation*}
and take the continuum limit.
Neglecting correlations between fluctuations of the operator quantum fields
$$\langle \hat{S}_a^{(n)} \hat{S}_b^{(n)}\rangle\,{\approx}\,
\langle \hat{S}_a^{(n)}\rangle \langle\hat{S}_b^{(n)}\rangle = \mu_a(x_n,t)\mu_b(x_n,t)$$
we obtain the following equation:
\begin{equation}\label{MFEqMot}
\hbar \frac{\partial \mu_a}{\partial t} = J \ell V_0\,  C_{abc} \mu_b \frac{\partial^2 \mu_c}{\partial x^2},
\end{equation}
where $\ell$ is the lattice distance, $V_0$ is an infinitesimal volume,
$C_{abc}$ are structure constants
of the algebra $\rmi\mathfrak{su}(3)$ with respect to the basis mentioned above.
It is easy to see that the equation is Hamiltonian and write the corresponding effective Hamiltonian:
\begin{equation*}
\hbar \frac{\partial \mu_a}{\partial t} = \{\hat{\mathcal{H}}_{\text{eff}},\mu_a\} =
V_0 C_{abc} \mu_b \frac{\delta \hat{\mathcal{H}}_{\text{eff}}}{\delta \mu_c},\qquad
\hat{\mathcal{H}}_{\text{eff}} = \frac{J \ell}{2} \int \sum_a \left(
\frac{\partial \mu_a}{\partial x}\right)^2 dx.
\end{equation*}
In the case of $d$-dimensional isotropic lattice the effective Hamiltonian has the form
\begin{equation*}
\hat{\mathcal{H}}_{\text{eff}} = \frac{J}{2\ell^{d-2}} \int \sum_a \left\langle
\frac{\partial \mu_a}{\partial \bm{x}},\frac{\partial \mu_a}{\partial \bm{x}}\right\rangle d \bm{x}.
\end{equation*}
For more details see \cite{BernHol2009}.

\subsection{Dynamical system of isotropic SU(3) magnet}
Here we construct a dynamical system
on a coadjoint orbit in the dual space $\widetilde{\mathfrak{g}}^{\ast}$ to the loop algebra
$\widetilde{\mathfrak{g}}\,{=}\,\rmi\mathfrak{su}(3)\,{\times}\, \mathcal{P}(\lambda,\lambda^{-1})$
with respect to the bilinear form
$\langle \hat{A}(\lambda), \hat{B}(\lambda) \rangle\,{=}\, \res_{\lambda=0}\lambda^{-N} \Tr \hat{A}(\lambda) \hat{B}(\lambda)$.
In $\rmi\mathfrak{su}(3)$ we introduce the same basis:
$\{\hat{S}_x$, $\hat{S}_y$, $\hat{S}_z$, $\hat{P}_{xy}$, $\hat{P}_{xz}$, $\hat{P}_{yz}$, $\hat{P}_{xx}\,{-}\,\hat{P}_{yy}$,
$\sqrt{3}(\hat{P}_{zz}\,{-}\,\frac{2}{3}\Ibb)\}$ and denote its elements by $\{\hat{S}_a\}$.
The loop algebra $\widetilde{\mathfrak{g}}$ is homogeneous and consists of the eigenspaces $\mathfrak{g}_m$, $m\,{\in}\,\Integer$ spanned over the basis elements $\{\lambda^m \hat{S}_a\}_{a=1}^8$. Evidently, $\widetilde{\mathfrak{g}}^{\ast}$
coincides with $\widetilde{\mathfrak{g}}$.

According to the Kostant-Adler scheme \cite{AdlerMoerb} $\widetilde{\mathfrak{g}}$ is decomposed into two subalgebras: $\widetilde{\mathfrak{g}}_+$ and $\widetilde{\mathfrak{g}}_-$ including the eigenspaces of nonnegative and negative degrees, respectively.
Consider the space $\mathcal{M}\,{\subset}\,\widetilde{\mathfrak{g}}^{\ast}$ of elements
\begin{multline}\label{LOp}
\hat{L} (\lambda) = \frac{\lambda^N}{2}\small\begin{pmatrix} m+q & 0 & 0 \\
 0 & -m+q & 0 \\ 0&0& -2q \end{pmatrix} + \frac{1}{2} \small \left(\begin{matrix}
 \mu_3(\lambda) + \tau_{33}(\lambda) \\
\tau_{11}(\lambda) - \tau_{22}(\lambda) + 2\rmi \tau_{12}(\lambda) \\
\frac{1}{\sqrt{2}}\,\big(\mu_1(\lambda)+2\rmi\tau_{23}(\lambda) + \rmi \mu_2(\lambda) + 2\tau_{13}(\lambda) \big)
\end{matrix} \right. \\  \small \left.
\begin{matrix}
\tau_{11}(\lambda) - \tau_{22}(\lambda) - 2\rmi \tau_{12}(\lambda) & \frac{1}{\sqrt{2}}\,
\big(\mu_1(\lambda)-2\rmi\tau_{23}(\lambda) - \rmi \mu_2(\lambda) + 2\tau_{13}(\lambda) \big) \\
-\mu_3(\lambda) + \tau_{33}(\lambda) &
\frac{1}{\sqrt{2}}\,\big(\mu_1(\lambda)-2\rmi\tau_{23}(\lambda) + \rmi \mu_2(\lambda) - 2\tau_{13}(\lambda) \big) \\
\frac{1}{\sqrt{2}}\,\big(\mu_1(\lambda)+2\rmi\tau_{23}(\lambda) - \rmi \mu_2(\lambda) - 2\tau_{13}(\lambda) \big) &
-2 \tau_{33}(\lambda)
\end{matrix} \right)
\end{multline}
with  constant $m$ and $q$. We denote the entries of $\hat{L}$ by $L_{ij}$, namely:
$L_{ij}(\lambda) \,{=}\, \sum_{m=0}^{N-1} L_{ij}^{(m)} \lambda^m$
and use the set $\{L_{ij}^{(m)}\}$, $m\,{=}\,0$, 1, \ldots, $N\,{-}\,1$ as dynamic variables in $\mathcal{M}$.
Obviously, $\mathcal{M}\,{\subset}\, (\widetilde{\mathfrak{g}}_+)^{\ast}$ with respect to the bilinear form, and
the factor-algebra $\widetilde{\mathfrak{g}}_+/\sum_{m\geqslant N} \mathfrak{g}_m$
acts effectively on $\mathcal{M}$. We call $\mathcal{M}$ the $N$-gap phase space.
In $\mathcal{M}$ we introduce the following Lie-Poisson bracket
\begin{gather}\label{LiePoissonBra}
  \{f_1,f_2\} = \sum_{m,n=0}^{N-1} \sum_{a,b=1}^{\rank \mathfrak{g}} P_{ab}^{mn}(N-1)
  \frac{\partial f_1}{\partial L_a^{(m)}}\frac{\partial
  f_2}{\partial L_b^{(n)}},  \\ P_{ab}^{mn}(N-1) = \langle \hat{L}(\lambda),
   [\hat{S}_a^{-m+N-1},\hat{S}_b^{-n+N-1}] \rangle,\qquad
   f_1,\, f_2 \in \mathcal{C}(\mathcal{M}), \notag
\end{gather}
giving the Poisson structure
\begin{equation*}
\{L_{ij}^{(m)},L_{kl}^{(n)}\} = L_{kj}^{(m+n+1-N)}\delta_{il} -
L_{il}^{(m+n+1-N)}\delta_{kj}.
\end{equation*}

Then we introduce the invariant functions $h_0$, $h_1$, \ldots, $h_{2N-1}$,
$f_0$, $f_1$, \ldots, $f_{3N-1}$ generated by
\begin{equation*}
I_2(\lambda) \equiv \tfrac{1}{2}\Tr \hat{L}^2 (\lambda) = \sum_{k=0}^{2N} h_k\lambda^{k}, \quad
I_3(\lambda) \equiv \tfrac{1}{3}\Tr \hat{L}^3 (\lambda) = \sum_{k=0}^{3N} f_k\lambda^{k}.
\end{equation*}
All these functions are in involution with respect to the Lie-Poisson bracket \eqref{LiePoissonBra},
so they are integrals of motion. Moreover, $h_0$, $h_1$, \ldots, $h_{N-1}$, $f_0$, $f_1$, \ldots, $f_{N-1}$
are functionally independent and annihilate the bracket, they define an orbit by means of the orbit equations
\begin{equation}\label{OrbitEq}
h_0=c_0,\quad h_1=c_1,\quad \dots,\quad h_{N-1}=c_{N-1},\quad f_0=b_0,\quad f_1=b_1,\quad \dots,\quad f_{N-1}=b_{N-1}.
\end{equation}
Other $3N$ integrals of motion serve as Hamiltonians for the dynamical system.

We recall the orbit structure of SU(3). It has two types of coadjoint orbits:
generic $\mathcal{O}_{\text{gen}}^{\text{SU}(3)}$ of dimension 6,
and degenerate $\mathcal{O}_{\text{deg}}^{\text{SU}(3)}$ of dimension 4:
\begin{equation*}
\mathcal{O}_{\text{gen}}^{\text{SU}(3)}
\,{=}\, \frac{\text{SU}(3)}{\text{U}(1)\times \text{U}(1)},\qquad
\mathcal{O}_{\text{deg}}^{\text{SU}(3)}
\,{=}\, \frac{\text{SU}(3)}{\text{SU}(2)\times \text{U}(1)}.
\end{equation*}
Every eigenspace $\mathfrak{g}_m\,{\subset}\,\mathcal{M}$, $m\,{=}\,0$, 1, \ldots, $N\,{-}\,1$,
is a foliation into such orbits under the coadjoint action of SU(3). An initial condition
defines what orbit is chosen in every $\mathfrak{g}_m$, and we call their direct product an orbit of the dynamical system.
So the phase space $\mathcal{M}$ is a foliation into the leaves which are direct products of $N$ samples of coadjoint orbits of SU(3). The system of isotropic SU(3) magnet lives in $\mathfrak{g}_0$, the evolution in all other eigenspaces is
defined by $\{L_{ij}^{(0)}\}$ and their derivatives with respect to $x$.

To proceed we introduce the spectral curve $\mathcal{R}$: 
$P(w,\lambda)\,{\equiv}\,\det \big(\hat{L}(\lambda) \,{-}\, w\Ibb\big)\,{=}\,0$.
In the generic case it is trigonal of genus $3N\,{-}\,2$:
\begin{equation}\label{SpectCurveEq}
w^3 - w\, I_2(\lambda) - I_3 (\lambda) = 0,
\end{equation}
the orbit equations are written explicitly \eqref{OrbitEq}.
In the degenerate case the spectral curve is decomposed as follows:
$$\big[w\mp 2\big(\tfrac{1}{3}\, I_2(\lambda)\big)^{1/2}\big]
\big[w\pm \big(\tfrac{1}{3}\, I_2(\lambda)\big)^{1/2}\big]^2 = 0$$
because of the relation $\big(\tfrac{1}{3}I_2(\lambda)\big)^3\,{=}\, \big({-}\tfrac{1}{2}I_3(\lambda)\big)^2$,
and is reduced to a hyperelliptic one.
The orbit equations are not so obvious as in the generic case.

The dynamical system in question is constructed from the Hamiltonian flows of $h_{N}$ and $h_{N+1}$:
\begin{equation*}
\begin{split}
&\frac{L_a^{(m)}}{\partial x} = \{L_a^{(m)},h_N\}\\
&\frac{L_a^{(m)}}{\partial t} = \{L_a^{(m)},h_{N+1}\}
\end{split}
\qquad \text{or} \qquad\quad
\begin{split}
&\frac{L(\lambda)}{\partial x} = [\nabla h_N,L(\lambda)]\\
&\frac{L(\lambda)}{\partial t} = [\nabla h_{N+1},L(\lambda)],
\end{split}
\end{equation*}
where $\nabla$ denotes the matrix gradient
\begin{equation*}
\nabla \mathcal{H} = \sum_{m=0}^{N-1} \sum_{a=1}^{\rank \mathfrak{g}}
\frac{\partial \mathcal{H}}{\partial L_a^{(m)}}\, \hat{Z}_a^{N-m-1},\qquad
 L_a^{(m)} = \langle \hat{L}(\lambda), \hat{Z}_a^{N-m-1}\rangle.
\end{equation*}
Commutativity of the Hamiltonian flows implies
the compatibility condition
\begin{gather}\label{ZeroCurvEq}
\frac{\partial \nabla h_{N}}{\partial t} -
\frac{\partial \nabla h_{N+1}}{\partial x} +
[\nabla h_{N},\nabla h_{N+1}] = 0\quad \Rightarrow \quad
\frac{\partial L_a^{(0)}}{\partial t} =
\frac{\partial L_a^{(1)}}{\partial x},
\end{gather}
which is the zero-curvature equation. We obtain different equations of motion over two types of orbits.
Over a degenerate orbit the equation \eqref{ZeroCurvEq} leads to
\begin{equation*}
\frac{\partial \mu_a}{\partial t} = \tfrac{2}{3h_0}  C_{abc} \mu_b \frac{\partial^2 \mu_c}{\partial x^2}
+ \tfrac{h_1}{2h_0}\, \frac{\partial \mu_a}{\partial x}
\end{equation*}
with $\mu_a(x,t) \,{\equiv}\, L_a^{(0)}(x,t)$.
The reader can recognize this equation as \eqref{MFEqMot}, obtained for the mean fields when correlations between
fluctuations are neglected.
Over a generic orbit the zero-curvature equation \eqref{ZeroCurvEq} acquires the more complicate form
\begin{multline}\label{EqMgen}
\frac{\partial \mu_a}{\partial t} = \tfrac{2}{4c_0^3-27 b_0^2}\,  C_{abc}
\bigg(c_0^2 \mu_b \frac{\partial^2 \mu_c}{\partial x^2}
- \tfrac{27}{2} b_0\Big(\mu_b \frac{\partial^2 \eta_c}{\partial x^2}
+ \eta_b \frac{\partial^2 \mu_c}{\partial x^2}\Big)
+  \tfrac{27}{4} c_0 \eta_b \frac{\partial^2 \eta_c}{\partial x^2} \bigg)
+ \\ + \tfrac{2 c_1 c_0^2 - 9b_1 b_0}{4c_0^3-27 b_0^2} \frac{\partial \mu_a}{\partial x}
+ \tfrac{9}{2}\,\tfrac{2b_1 c_0 - 3c_1 b_0}{4c_0^3-27 b_0^2} \frac{\partial \eta_a}{\partial x}.
\end{multline}
The latter also serves as equation of motion with the effective Hamiltonian ($c_1\,{=}\,b_1\,{=}\,0$)
\begin{equation*}
\hat{\mathcal{H}}_{\text{eff}} = \tfrac{1}{4c_0^3-27 b_0^2} \int \sum_a \left(
c_0^2 \bigg(\frac{\partial \mu_a}{\partial x}\bigg)^2\!\! - 27 b_0
\frac{\partial \mu_a}{\partial x}\frac{\partial \eta_a}{\partial x} +
\tfrac{27}{4} c_0 \bigg(\frac{\partial \eta_a}{\partial x}\bigg)^2 \right)\! dx,
\end{equation*}
where $\eta_a \,{=}\, d_{abc} \mu_b \mu_c$ with the symmetric tensor
$d_{abc}\,{=}\,\Tr (\hat{S}_a \hat{S}_b \hat{S}_c \,{+}\, \hat{S}_b \hat{S}_a \hat{S}_c)$.
This dynamical system was firstly presented in \cite{SaRuij1982}, and constructed by means of the orbit method
in \cite{BernHol2009}.

\section{Integration of the system over a generic orbit}
\subsection{Separation of variables}
We start the process of solving \eqref{EqMgen} from separation of variables.
In \cite{Sklya95} Sklyanin declared
\begin{conjecture*}
Suppose the orbit $\mathcal{O}_{\text{gen}}^{\text{SU(3)}}$ has the coordinates
 $\{L_{11}^{(m)}$, $L_{12}^{(m)}$, $L_{21}^{(m)}$, $L_{22}^{(m)}$, $L_{31}^{(m)}$,
 $L_{32}^{(m)}\,{:}$ $m\,{=}\,0,\,\dots,\,N\,{-}\,1\}$ as above.
Then the new coordinates $\{(\lambda_k, w_k)\,{:}$ $k\,{=}\,1,\,\dots,\,3N\}$,
defined by the formulas
\begin{equation*}
  \mathcal{B}(\lambda_k)=0,\qquad
  w_k = \mathcal{A}(\lambda_k),
\end{equation*}
where
\begin{align*}
&\mathcal{B}(\lambda)= \big[L_{11}(\lambda) - L_{22}(\lambda) \big] L_{31}(\lambda)L_{32}(\lambda)-
L_{12}(\lambda) L_{31}^2(\lambda) + L_{21}(\lambda) L_{32}^2(\lambda),\\
&\mathcal{A}(\lambda_k) = L_{11}(\lambda_k) -\frac{L_{12}(\lambda_k) L_{31}(\lambda_k)}{L_{32}(\lambda_k)}
 = L_{22}(\lambda_k) - \frac{L_{21}(\lambda_k) L_{32}(\lambda_k)}{L_{31}(\lambda_k)}
\end{align*}
serve as variables of separation.
\end{conjecture*}
The polynomial $\mathcal{B}$ of degree $3N$ gives a half of variables of separation,
we denote them by $\{\lambda_k\}$. The function $\mathcal{A}$
gives the other half of separation variables, denoted by $\{w_k\}$.
The same expressions for $\mathcal{B}$ and $\mathcal{A}$ were used by
Sklyanin for the SL(3) magnetic chain model with a quadratic Poisson bracket.

Separation of variables was developed in many papers, among them \cite{Adams, Scott, Gekhtman, Sklya95}.
Using their ideas it is easy to prove the following statement for our system.
\begin{proposition*}
The variables of separation have the following properties:
\begin{enumerate}
\item[\textup{(1)}]  a pair $(w_k,\lambda_k)$ is a point of the spectral curve \eqref{SpectCurveEq};
\item[\textup{(2)}] a pair
$(w_k, \lambda_k)$ is quasi-canonically conjugate with respect to
the Lie-Poisson bracket \eqref{LiePoissonBra}:
\begin{equation*}
  \{\lambda_k,\lambda_l\} =0, \qquad
  \{\lambda_k, w_l\} = -\lambda_k^{N}\delta_{kl}, \qquad
  \{w_k,w_l\}=0;
\end{equation*}
\item[\textup{(3)}] the corresponding  Liouville 1-form is
$\Omega= - \sum\limits_{k} \lambda_k^{-N} w_{k}\,d\lambda_{k}.$
\end{enumerate}
\end{proposition*}

We realize this separation of variables for the system of isotropic SU(3) magnet over a generic orbit.
In other words, we solve the problem how to express dynamic variables in terms of the variables of separation.
We start from the dynamic variables, which are coefficients of entries of the $L$-matrix \eqref{LOp},
and construct a sequence of maps leading to variables of separation. These maps are evidently invertible.

First of all we need to eliminate some dynamic variables from the set
$\{L_{ij}^{(m)},\,L_{11}^{(m)}+L_{22}^{(m)}+L_{33}^{(m)}\,{=}\,0\,{:}$
$m\,{=}\,0,\,\dots,\,N\,{-}\,1\}$ in order to parameterize the orbit.
There are $8N$ dynamic variables on a $6N$-dimensional orbit, so we need to eliminate $2N$ variables:
one can choose the variables corresponding to a pair of commuting nilpotent elements of the Cartan-Weyl basis,
that is one of the following sets: $\{L_{13}^{(m)},\,L_{23}^{(m)}\}$, $\{L_{12}^{(m)},\,L_{32}^{(m)}\}$, $\{L_{21}^{(m)},\,L_{31}^{(m)}\}$, $\{L_{12}^{(m)},\,L_{13}^{(m)}\}$, $\{L_{21}^{(m)},\,L_{23}^{(m)}\}$, $\{L_{31}^{(m)},\,L_{32}^{(m)}\}$.
The orbit equations \eqref{OrbitEq} are linear in any of these sets of dynamic variables.

For the eliminated set $\{L_{13}^{(m)},\,L_{23}^{(m)}\}$ the orbit equations are written as follows:
\begin{gather*}
\small
\begin{pmatrix} c_0\vphantom{L_{31}^{(0)}} \\ b_0\vphantom{L_{31}^{(0)}} \\
c_1\vphantom{L_{31}^{(0)}} \\ b_1\vphantom{L_{31}^{(0)}} \\ \vdots \\
c_{N-1}\vphantom{L_{31}^{(0)}} \\ b_{N-1}\vphantom{L_{31}^{(0)}} \end{pmatrix} =
\begin{pmatrix} L_{31}^{(0)} & L_{32}^{(0)} & 0 & 0 & \dots & 0 & 0 \\
\mathcal{K}_{31}^{(0)} & \mathcal{K}_{32}^{(0)} & 0 & 0 & \dots & 0 & 0 \\
L_{31}^{(1)} & L_{32}^{(1)} & L_{31}^{(0)} & L_{32}^{(0)} & \dots & 0 & 0 \\
\mathcal{K}_{31}^{(1)} & \mathcal{K}_{32}^{(1)} & \mathcal{K}_{31}^{(0)} & \mathcal{K}_{32}^{(0)} & \dots & 0 & 0 \\
\vdots & \vdots & \vdots & \vdots & \ddots & \vdots & \vdots \\
L_{31}^{(N-1)} & L_{32}^{(N-1)} & L_{31}^{(N-2)} & L_{32}^{(N-2)} & \dots & L_{31}^{(0)} & L_{32}^{(0)} \\
\mathcal{K}_{31}^{(N-1)} & \mathcal{K}_{32}^{(N-1)} & \mathcal{K}_{31}^{(N-2)} & \mathcal{K}_{32}^{(N-2)}
& \dots & \mathcal{K}_{31}^{(0)} & \mathcal{K}_{32}^{(0)} \end{pmatrix}
\begin{pmatrix} L_{13}^{(0)} \\ L_{23}^{(0)} \\ L_{13}^{(1)} \\ L_{23}^{(1)} \\
\vdots \\ L_{13}^{(N-1)} \\ L_{23}^{(N-1)} \end{pmatrix} +
\begin{pmatrix} H_0\vphantom{L_{31}^{(0)}} \\ F_0\vphantom{L_{31}^{(0)}} \\
H_1\vphantom{L_{31}^{(0)}} \\ F_1\vphantom{L_{31}^{(0)}} \\ \vdots \\
H_{N-1}\vphantom{L_{31}^{(0)}} \\ F_{N-1}\vphantom{L_{31}^{(0)}}  \end{pmatrix},\\
H_j = - \mathcal{K}_{33}^{(j)} + \sum\limits_{m+n=j} L_{33}^{(m)}L_{33}^{(n)}, \qquad
F_j = \sum\limits_{m+n=j} L_{33}^{(m)} \mathcal{K}_{33}^{(n)},
\end{gather*}
where $\mathcal{K}_{ij}$ are cofactors of $L_{ji}$.
Solving this linear problem we express the variables $\{L_{13}^{(m)},\,L_{23}^{(m)}\}$ in terms of other dynamic variables,
then we eliminate them from the integrals of motion $h_N$, $h_{N+1}$, \ldots, $h_{2N-1}$,
$f_N$, $f_{N+1}$, \ldots, $f_{3N-1}$, serving as Hamiltonians:
\begin{gather}\label{HamDynVar}
\scriptsize
\begin{pmatrix} h_N\vphantom{L_{31}^{(0)}} \\ f_N\vphantom{L_{31}^{(0)}} \\
h_{N+1}\vphantom{L_{31}^{(0)}} \\ f_{N+1}\vphantom{L_{31}^{(0)}} \\
\vdots \\ h_{2N-1}\vphantom{L_{31}^{(0)}} \\ f_{2N-1}\vphantom{L_{31}^{(0)}} \\
f_{2N}\vphantom{L_{31}^{(0)}} \\
f_{2N+1}\vphantom{L_{31}^{(0)}} \\ \vdots \\
f_{3N-1}\vphantom{L_{31}^{(0)}} \end{pmatrix} =
\begin{pmatrix} L_{31}^{(N)} & L_{32}^{(N)} & L_{31}^{(N-1)} & L_{32}^{(N-1)} & \dots
& L_{31}^{(1)} & L_{32}^{(1)} \\
\mathcal{K}_{31}^{(N)} & \mathcal{K}_{32}^{(N)} & \mathcal{K}_{31}^{(N-1)} & \mathcal{K}_{32}^{(N-1)} & \dots
& \mathcal{K}_{31}^{(1)} & \mathcal{K}_{32}^{(1)} \\
0 & 0 & L_{31}^{(N)} & L_{32}^{(N)} & \dots
& L_{31}^{(2)} & L_{32}^{(2)} \\
\mathcal{K}_{31}^{(N+1)} & \mathcal{K}_{32}^{(N+1)} & \mathcal{K}_{31}^{(N)} & \mathcal{K}_{32}^{(N)} & \dots
& \mathcal{K}_{31}^{(2)} & \mathcal{K}_{32}^{(2)} \\
\vdots & \vdots & \vdots & \vdots & \ddots & \vdots & \vdots \\
0 & 0 & 0 & 0 & \dots & L_{31}^{(N)} & L_{32}^{(N)} \\
\mathcal{K}_{31}^{(2N-1)} & \mathcal{K}_{32}^{(2N-1)} &
\mathcal{K}_{31}^{(2N-2)} & \mathcal{K}_{32}^{(2N-2)} & \dots &
\mathcal{K}_{31}^{(N)} & \mathcal{K}_{32}^{(N)} \\
\mathcal{K}_{31}^{(2N)} & \mathcal{K}_{32}^{(2N)} &
\mathcal{K}_{31}^{(2N-1)} & \mathcal{K}_{32}^{(2N-1)} & \dots &
\mathcal{K}_{31}^{(N+1)} & \mathcal{K}_{32}^{(N+1)} \\
0 & 0 & \mathcal{K}_{31}^{(2N)} & \mathcal{K}_{32}^{(2N)} & \dots &
\mathcal{K}_{31}^{(N+2)} & \mathcal{K}_{32}^{(N+2)} \\
\vdots & \vdots & \vdots & \vdots & \ddots & \vdots & \vdots \\
0 & 0 & 0 & 0 & \dots & \mathcal{K}_{31}^{(2N)} & \mathcal{K}_{32}^{(2N)} \end{pmatrix}
\begin{pmatrix} L_{13}^{(0)}  \\ L_{23}^{(0)} \\ L_{13}^{(1)} \\ L_{23}^{(1)} \\
\vdots \\ L_{13}^{(N-1)} \\ L_{23}^{(N-1)} \end{pmatrix} +
\begin{pmatrix} H_N\vphantom{L_{31}^{(0)}} \\ F_N\vphantom{L_{31}^{(0)}} \\
H_{N+1}\vphantom{L_{31}^{(0)}} \\ F_{N+1}\vphantom{L_{31}^{(0)}} \\ \vdots \\
H_{2N-1}\vphantom{L_{31}^{(0)}} \\ F_{2N-1}\vphantom{L_{31}^{(0)}}
\\ F_{2N}\vphantom{L_{31}^{(0)}} \\ F_{2N+1}\vphantom{L_{31}^{(0)}} \\ \vdots
\\ F_{3N-1}\vphantom{L_{31}^{(0)}} \end{pmatrix}.
\end{gather}

On the other hand the Hamiltonians $h_N$, $h_{N+1}$, \ldots, $h_{2N-1}$,
$f_N$, $f_{N+1}$, \ldots, $f_{3N-1}$ can be computed from the spectral curve equation
written for $3N$ points $(w_k,\lambda_k)$:
\begin{multline}\label{HyperCurveHM}
  w_k^3 = w_k\Big(c_{0}+c_1 \lambda_k + \cdots +
  c_{N-1} \lambda_k^{N-1} + h_{N} \lambda_k^{N} + \cdots + h_{2N-1} \lambda_k^{2N-1}
  + c_{2N} \lambda^{2N}\Big)
  + \\ +\Big(b_{0} + b_1 \lambda_k + \cdots + b_{N-1}\lambda_k^{N-1} +
  f_{N} \lambda_k^{N} + \cdots + f_{3N-1} \lambda_k^{3N-1}
  + b_{3N} \lambda_k^{3N}\Big),
\end{multline}
where $c_{2N}\,{=}\,\frac{1}{4}(m^2\,{+}\,3q^2)$, $b_{3N}\,{=}\,\frac{1}{4} q(m^2\,{-}\,q^2)$.
Equating expressions \eqref{HamDynVar} for the Hamiltonians
to the obtained from \eqref{HyperCurveHM} we get the equations
\begin{gather}\label{SepVarEq}
L_{31}(\lambda_k) w_k + \mathcal{K}_{31}(\lambda_k)=0,\qquad
L_{32}(\lambda_k) w_k + \mathcal{K}_{32}(\lambda_k)=0.
\end{gather}
giving the function $\mathcal{A}$.
The consistent equation with respect to $\{w_k\}$ gives the polynomial $\mathcal{B}$.

The equations \eqref{SepVarEq} are linear in the  set $\{L_{31}^{(0)}$, $L_{32}^{(0)}$, \ldots, $L_{31}^{(N-1)}$, $L_{32}^{(N-1)}$, $\mathcal{K}_{31}^{(0)}$, $\mathcal{K}_{32}^{(0)}$, \ldots, $\mathcal{K}_{31}^{(2N-1)}$, $\mathcal{K}_{32}^{(2N-1)}\}$, that is
\begin{gather*}
\widehat{W} \bm{L}_1 = - L_{31}^{(N)} \bm{w} - \mathcal{K}_{31}^{(2N)} \bm{\lambda},
\qquad
\widehat{W} \bm{L}_2 = - L_{32}^{(N)} \bm{w} - \mathcal{K}_{32}^{(2N)} \bm{\lambda},\\
\widehat{W} = \small \begin{bmatrix}
\lambda_1^{N-1} w_1 & \dots & \lambda_1 w_1 & w_1 & \lambda_1^{2N-1}
& \dots & \lambda_1 & 1 \\
\lambda_2^{N-1} w_2 & \dots & \lambda_2 w_2 & w_2 & \lambda_2^{2N-1}
& \dots & \lambda_2 & 1 \\
\vdots & \ddots & \vdots & \vdots & \vdots & \ddots & \vdots & \vdots \\
\lambda_{3N}^{N-1} w_{3N} & \dots & \lambda_{3N} w_{3N} & w_{3N} & \lambda_{3N}^{2N-1}
& \dots & \lambda_{3N} & 1 \end{bmatrix},\ \
\bm{w} = \begin{pmatrix} \lambda_1^{N} w_1 \\
\lambda_2^{N} w_2 \\ \vdots \\ \lambda_{3N}^{N} w_{3N} \end{pmatrix},\ \
\bm{\lambda} = \begin{pmatrix} \lambda_1^{2N} \\
\lambda_2^{2N} \\ \vdots \\ \lambda_{3N}^{2N}  \end{pmatrix},\\
\bm{L}_1^{t} = \begin{bmatrix} L_{31}^{(N-1)}, & \dots, & L_{31}^{(1)}, & L_{31}^{(0)}, &
\mathcal{K}_{31}^{(2N-1)}, & \dots, & \mathcal{K}_{31}^{(1)}, & \mathcal{K}_{31}^{(0)} \end{bmatrix},\\
\bm{L}_2^{t} = \begin{bmatrix} L_{32}^{(N-1)}, & \dots, & L_{32}^{(1)}, & L_{32}^{(0)}, &
\mathcal{K}_{32}^{(2N-1)}, & \dots, & \mathcal{K}_{32}^{(1)}, & \mathcal{K}_{32}^{(0)} \end{bmatrix},
\end{gather*}
and can be easily solved for this set of dynamic variables.
The next step is obtaining
$\{L_{11}^{(m)}$, $L_{12}^{(m)}$, $L_{21}^{(m)}$, $L_{22}^{(m)}\,{:}$ $m\,{=}\,1,$ \ldots, $N\,{-}\,1\}$
from the expressions for $\{\mathcal{K}_{31}^{(m)}$, $\mathcal{K}_{32}^{(m)}\,{:}\;m\,{=}\,1,\,\ldots,\,2N\,{-}\,1\}$,
which is also a linear problem:
\begin{gather*}
\small
\begin{pmatrix}
\mathcal{K}_{31}^{(0)} \\ \mathcal{K}_{31}^{(1)} \\ \vdots \\ \mathcal{K}_{31}^{(N-1)} \\
\mathcal{K}_{31}^{(N)} \\ \mathcal{K}_{31}^{(N+1)} \\ \vdots \\ \mathcal{K}_{31}^{(2N-1)}
\end{pmatrix} =
\begin{pmatrix}
\begin{matrix} L_{32}^{(0)} & 0 & \dots & 0  \\
 L_{32}^{(1)} & L_{32}^{(0)} & \dots & 0 \\
 \vdots & \vdots & \ddots & \vdots  \\
 L_ {32}^{(N-1)} & L_{32}^{(N-2)} & \dots & L_{32}^{(0)}\\
 L_{32}^{(N)} & L_{32}^{(N-1)} & \dots & L_{32}^{(1)} \\
 0 & L_{32}^{(N)} & \dots & L_{32}^{(2)} \\
 \vdots & \vdots & \ddots & \vdots \\
 0 & 0 & \dots & L_{32}^{(N)}
\end{matrix} &
\begin{matrix} -L_{31}^{(0)} & 0 & \dots & 0  \\
 -L_{31}^{(1)} & -L_{31}^{(0)} & \dots & 0 \\
 \vdots & \vdots & \ddots & \vdots  \\
 -L_ {31}^{(N-1)} & -L_{31}^{(N-2)} & \dots & -L_{31}^{(0)}\\
 -L_{31}^{(N)} & -L_{31}^{(N-1)} & \dots & -L_{31}^{(1)} \\
 0 & -L_{31}^{(N)} & \dots & -L_{31}^{(2)} \\
 \vdots & \vdots & \ddots & \vdots \\
 0 & 0 & \dots & -L_{31}^{(N)}
\end{matrix}
\end{pmatrix}
\begin{pmatrix}
L_{21}^{(0)} \\ L_{21}^{(1)} \\ \vdots \\ L_{21}^{(N-1)} \\
L_{22}^{(0)} \\ L_{22}^{(1)} \\ \vdots \\ L_{22}^{(N-1)}
\end{pmatrix},\\
\small
\begin{pmatrix}
\mathcal{K}_{32}^{(0)} \\ \mathcal{K}_{32}^{(1)} \\ \vdots \\ \mathcal{K}_{32}^{(N-1)} \\
\mathcal{K}_{32}^{(N)} \\ \mathcal{K}_{32}^{(N+1)} \\ \vdots \\ \mathcal{K}_{32}^{(2N-1)}
\end{pmatrix} =
\begin{pmatrix}
\begin{matrix} -L_{32}^{(0)} & 0 & \dots & 0  \\
 -L_{32}^{(1)} & -L_{32}^{(0)} & \dots & 0 \\
 \vdots & \vdots & \ddots & \vdots  \\
 -L_ {32}^{(N-1)} & -L_{32}^{(N-2)} & \dots & -L_{32}^{(0)}\\
 -L_{32}^{(N)} & -L_{32}^{(N-1)} & \dots & -L_{32}^{(1)} \\
 0 & -L_{32}^{(N)} & \dots & -L_{32}^{(2)} \\
 \vdots & \vdots & \ddots & \vdots \\
 0 & 0 & \dots & -L_{32}^{(N)}
\end{matrix} &
\begin{matrix} L_{31}^{(0)} & 0 & \dots & 0  \\
 L_{31}^{(1)} & L_{31}^{(0)} & \dots & 0 \\
 \vdots & \vdots & \ddots & \vdots  \\
 L_ {31}^{(N-1)} & L_{31}^{(N-2)} & \dots & L_{31}^{(0)}\\
 L_{31}^{(N)} & L_{31}^{(N-1)} & \dots & L_{31}^{(1)} \\
 0 & L_{31}^{(N)} & \dots & L_{31}^{(2)} \\
 \vdots & \vdots & \ddots & \vdots \\
 0 & 0 & \dots & L_{31}^{(N)}
\end{matrix}
\end{pmatrix}
\begin{pmatrix}
L_{11}^{(0)} \\ L_{11}^{(1)} \\ \vdots \\ L_{11}^{(N-1)} \\
L_{12}^{(0)} \\ L_{12}^{(1)} \\ \vdots \\ L_{12}^{(N-1)}
\end{pmatrix}.
\end{gather*}
The rest of variables, eliminated at the first stage, we compute from the orbit equations.

\begin{note*}
The proposed scheme fails in application to the $L$-matrix \eqref{LOp}, because degrees of
the polynomials $\mathcal{B}$, $\mathcal{K}_{31}$, $\mathcal{K}_{32}$
are less than the maximal. We avoid this failure if rotate the $L$-matrix by means
of similar transformation with the matrix
\begin{equation*}
\small \hat{R} = \begin{pmatrix}
\frac{1}{\sqrt{2}} & \frac{1}{2} & \frac{1}{2}\\
-\frac{1}{\sqrt{2}} & \frac{1}{2} & \frac{1}{2}\\
0 & -\frac{1}{\sqrt{2}} & \frac{1}{\sqrt{2}}
\end{pmatrix},\qquad
\hat{R}^{-1}\hat{L} (\lambda) \hat{R} = \frac{\lambda^N}{2}
\begin{pmatrix}
q & \frac{m}{\sqrt{2}} & \frac{m}{\sqrt{2}} \\
\frac{m}{\sqrt{2}} & -\frac{q}{2} & \frac{3q}{2} \\
\frac{m}{\sqrt{2}} & \frac{3q}{2} & -\frac{q}{2} \end{pmatrix}
+ \cdots
\end{equation*}
\end{note*}

\subsection{Finite-gap integration}
As mentioned above the Hamiltonians $h_N$, $h_{N+1}$ give rise to a stationary and an evolutionary flows, 
resulting in the equations
\begin{align*}
&\frac{d\lambda_k}{dx} = \frac{\big[L_{32}^{(0)} L_{31}(\lambda_k) - L_{31}^{(0)} L_{32}(\lambda_k) \big]
\frac{\partial}{\partial w_k} P(w_k,\lambda_k)}{\mathcal{B}_{3N} \lambda_k \prod'_j (\lambda_k-\lambda_j)},\\
&\frac{d\lambda_k}{dt} = \frac{\big[\big(L_{32}^{(0)} + L_{32}^{(1)} \lambda_k\big) L_{31}(\lambda_k)
- \big(L_{31}^{(0)} + L_{31}^{(1)} \lambda_k\big) L_{32}(\lambda_k) \big]
\frac{\partial}{\partial w_k} P(w_k,\lambda_k)}
{\mathcal{B}_{3N} \lambda_k^2 \prod'_j (\lambda_k-\lambda_j)}
\end{align*}
for $3N$ points $(w_k,\lambda_k)$ on the spectral curve $\mathcal{R}$ of genus $3N\,{-}\,2$. 
We introduce the basis of holomorphic Abelian differentials on the curve:
\begin{equation*}
\frac{\lambda^n\, d\lambda}{\frac{\partial}{\partial w} P(w,\lambda)},\quad n\,{=}\,0,\,1,\,\dots,\,2N-2,\quad
\frac{w \lambda^n\, d\lambda}{\frac{\partial}{\partial w} P(w,\lambda)},\quad n\,{=}\,0,\,1,\,\dots,\,N-2,
\end{equation*}
where zeros of $\frac{\partial}{\partial w} P(w,\lambda)\,{=}\,3w^2 \,{-}\, I_2(\lambda)$ belong to the set of zeros
of the discriminant $Q$ of the trigonal curve \eqref{SpectCurveEq}: $Q(\lambda) \,{=}\, \big(\tfrac{1}{3}\, I_2(\lambda)\big)^3 \,{-}\, \big(\tfrac{1}{2}I_3(\lambda)\big)^2$, giving branch points to the curve. 
Also we need two Abelian differentials of the third kind:
\begin{equation*}
 \frac{\lambda^{2N-1}\,d\lambda }{\frac{\partial}{\partial w} P(w,\lambda)},\qquad
 \frac{w\lambda^{N-1}\,d\lambda }{\frac{\partial}{\partial w} P(w,\lambda)}.
\end{equation*}

In general a finite gap integration of similar systems was done in \cite{Adams}. In more details
integration in terms of generalized $\theta$-functions giving inversion of an extended Abel-Jacobi map
can be found in \cite{Fed1999,BraFed2008}.

Here we consider the one-gap system having only the stationary flow,
resulting in the following equations for the variables $\{\lambda_k\}$:
\begin{gather*}
\begin{split}
&\frac{\partial \lambda_k}{\partial x} =
-\frac{\big[ 3w_1^2 - I_2(\lambda_k)\big](\lambda_i-\lambda_j)}{\det \widehat{W}},
\end{split}
\end{gather*}
where $(k\,ij)$ is a cyclic permutation of (123).
The map $(\lambda_1,\lambda_2,\lambda_3)\mapsto
(\varphi_1,\varphi_2,\varphi_3)$ is nonsingular if $\det \widehat{W}\,{\neq}\,0$, and leads to the Abelian integrals
\begin{gather*}
\varphi_1 = \int^{\lambda_1}_{\lambda_0} \frac{d\lambda }{3w^2 - I_2(\lambda)}+
\int^{\lambda_2}_{\lambda_0} \frac{d\lambda }{3w^2 - I_2(\lambda)} +
\int^{\lambda_3}_{\lambda_0} \frac{d\lambda }{3w^2 - I_2(\lambda)},\\
\varphi_2 = \int^{\lambda_1}_{\lambda_0} \frac{\lambda d\lambda }{3w^2 - I_2(\lambda)}+
\int^{\lambda_2}_{\lambda_0} \frac{\lambda d\lambda }{3w^2 - I_2(\lambda)} +
\int^{\lambda_3}_{\lambda_0} \frac{\lambda d\lambda }{3w^2 - I_2(\lambda)},\\
\varphi_3 = \int^{\lambda_1}_{\lambda_0} \frac{w\,d\lambda }{3w^2 - I_2(\lambda)}+
\int^{\lambda_2}_{\lambda_0} \frac{w\,d\lambda }{3w^2 - I_2(\lambda)} +
\int^{\lambda_3}_{\lambda_0} \frac{w\,d\lambda }{3w^2 - I_2(\lambda)};\\
\frac{d\varphi_1}{dx} = 0,\quad \frac{d\varphi_2}{dx} = 0,\quad
\frac{d\varphi_3}{dx} = -1\quad \Rightarrow \quad
\varphi_1 = C_1,\quad \varphi_2 = C_2,\quad
\varphi_3 = C_3 - x.
\end{gather*}
The first differential is holomorphic, and two others
are meromorphic of the third kind.

The spectral curve of the one-gap system is elliptic. We reduce it to
the Weierstrass form by rotating about an inflection point $(\lambda_{\text{f}},w_{\text{f}})$, which is a solution of the equation
\begin{gather*}
\big[3 w_{\text{f}}^2 + I_2(\lambda_{\text{f}})  \big]\big[\tfrac{3}{2} I''_2(\lambda_{\text{f}}) I_3(\lambda_{\text{f}})+ \tfrac{1}{2} I''_3(\lambda_{\text{f}}) I_2(\lambda_{\text{f}})- I_2'(\lambda_{\text{f}}) I_3'(\lambda_{\text{f}})\big] + \\
+ w_{\text{f}} \big[2 I''_2(\lambda_{\text{f}}) I_2^2(\lambda_{\text{f}})+\tfrac{9}{2}  I''_3(\lambda_{\text{f}}) I_3(\lambda_{\text{f}})
- I_2(\lambda_{\text{f}})[I_2'(\lambda_{\text{f}})]^2 - 3[I_3'(\lambda_{\text{f}})]^2\big]=0.
\end{gather*}
The rotation by means of a linear transformation together with a linear-fractional transformation
realizes the map $(w,\lambda)\mapsto (z,\xi)$:
\begin{gather}\label{WPmap}
\begin{pmatrix} \lambda - \lambda_{\text{f}} \\ w-w_{\text{f}} \end{pmatrix} =
\small \begin{pmatrix}  3 w_{\text{f}}^2 - I_2(\lambda_{\text{f}})&
-\frac{1}{2}\,\big[w_{\text{f}} \, I_2'(\lambda_{\text{f}}) + I_3'(\lambda_{\text{f}})\big]^{-1} \\
w_{\text{f}} \, I_2'(\lambda_{\text{f}}) + I_3'(\lambda_{\text{f}}) &
\frac{1}{2}\,\big[ 3 w_{\text{f}}^2 - I_2(\lambda_{\text{f}})\big]^{-1} \end{pmatrix}
\frac{2}{z-d+b(\xi - a)}
\begin{pmatrix} -(\xi - a) \\ c \end{pmatrix},
\end{gather}
where
\begin{align*}
&a = \tfrac{1}{12}\bigg[\frac{\big(6w_{\text{f}} [w_{\text{f}} \, I_2'(\lambda_{\text{f}}) + I_3'(\lambda_{\text{f}})]
-I'_2(\lambda_{\text{f}})[3 w_{\text{f}}^2 - I_2(\lambda_{\text{f}})]\big)^2}{[3 w_{\text{f}}^2 - I_2(\lambda_{\text{f}})]^2} - \\
&-4\frac{\big(3[w_{\text{f}}  I_2'(\lambda_{\text{f}}) + I_3'(\lambda_{\text{f}})]^3
+h_2[3 w_{\text{f}}^2 - I_2(\lambda_{\text{f}})]^2 [w_{\text{f}}  I_2'(\lambda_{\text{f}}) + I_3'(\lambda_{\text{f}})] + 3f_3[3 w_{\text{f}}^2 - I_2(\lambda_{\text{f}})]^3\big)I'_2(\lambda_{\text{f}})}
{2[3 w_{\text{f}}^2 - I_2(\lambda_{\text{f}})][w_{\text{f}}  I_2'(\lambda_{\text{f}}) + I_3'(\lambda_{\text{f}})]}\bigg],\\
&b = \frac{6w_{\text{f}} [w_{\text{f}} \, I_2'(\lambda_{\text{f}}) + I_3'(\lambda_{\text{f}})]-I'_2(\lambda_{\text{f}})[3 w_{\text{f}}^2 - I_2(\lambda_{\text{f}})]}
{[3 w_{\text{f}}^2 - I_2(\lambda_{\text{f}})]},\\
&c = [w_{\text{f}}  I_2'(\lambda_{\text{f}}) + I_3'(\lambda_{\text{f}})]^3
-h_2[3 w_{\text{f}}^2 - I_2(\lambda_{\text{f}})]^2 [w_{\text{f}}  I_2'(\lambda_{\text{f}}) + I_3'(\lambda_{\text{f}})] - f_3[3 w_{\text{f}}^2 - I_2(\lambda_{\text{f}})]^3,\\
&d= \frac{\big([w_{\text{f}}  I_2'(\lambda_{\text{f}}) + I_3'(\lambda_{\text{f}})]^3
-h_2[3 w_{\text{f}}^2 - I_2(\lambda_{\text{f}})]^2 [w_{\text{f}}  I_2'(\lambda_{\text{f}}) + I_3'(\lambda_{\text{f}})] - f_3[3 w_{\text{f}}^2 - I_2(\lambda_{\text{f}})]^3\big)I'_2(\lambda_{\text{f}})}{2[3 w_{\text{f}}^2 - I_2(\lambda_{\text{f}})][w_{\text{f}}  I_2'(\lambda_{\text{f}}) + I_3'(\lambda_{\text{f}})]}.
\end{align*}
The map \eqref{WPmap} reduces the spectral curve \eqref{SpectCurveEq} to the Weierstrass form
\begin{gather*}
z^2 = 4 \xi^3 - g_2 \xi - g_3.
\end{gather*}
Then the parametrization $\xi\,{=}\,\wp(u)$, $z\,{=}\,\wp'(u)$ makes a uniformization of the curve.
In the new variables the Abelian integrals are reduced to the form
\begin{gather*}
\varphi_1 = \sum_{k=1}^3 \int^{u_k}_{u_0} du,\qquad 
\varphi_2 = \sum_{k=1}^3 \int^{u_k}_{u_0} \bigg(\lambda_{\text{f}} - \frac{2[\wp(u) - a]}{\wp'(u)-d+b[\wp(u) - a]}\bigg) du,\\
\varphi_3 = \sum_{k=1}^3 \int^{u_k}_{u_0} \bigg(w_{\text{f}} + \frac{2c}{\wp'(u)-d+b[\wp(u) - a]}\bigg)du.
\end{gather*}


\ack
This work is supported by the International Charitable Fund for Renaissance of
Kyiv-Mohyla Academy. We thank the organizing committee of the conference `Physics and Mathematics
of Nonlinear Phenomena 2013' for the substantial financial support and warm hospitality.

\section*{References}

\end{document}